\documentclass{agujournal2019}

\usepackage{xcolor}
\usepackage{lscape}
\usepackage{longtable}
\usepackage{array}
\usepackage{amsmath}
\usepackage{amssymb}
\usepackage{multirow}
\usepackage{soul}
\usepackage{lineno}
\usepackage{mathtools}
\usepackage{float}
\usepackage{comment}
\begin{document}

% \begin{frontmatter}

% \title{Thermal experiments for fractured rock characterization: theoretical analysis and inverse modeling}

% \author[label1]{Zitong Zhou}
% \author[label2]{Delphine Roubinet\corref{cor2}}
% \author[label1]{Daniel M. Tartakovsky\corref{cor1} }

% \address[label1]{Department of Energy Resources Engineering, Stanford University, Stanford, CA 94305, USA}
% \address[label2]{Geosciences Montpellier (UMR 5243),
% University of Montpellier, 4090 Montpellier, France}

% \cortext[cor1]{Corresponding author. \textit{Email address}: tartakovsky@stanford.edu}
% \cortext[cor2]{Corresponding author. \textit{Email address}: delphine.roubinet@umontpellier.fr}
\title{Thermal experiments for fractured rock characterization: theoretical analysis and inverse modeling}
\authors{Zitong Zhou\affil{1}, Delphine Roubinet\affil{2}, and Daniel M. Tartakovsky \affil{1}}

\affiliation{1}{Department of Energy Resources Engineering, Stanford University, Stanford, CA 94305, USA}
\affiliation{2}{Geosciences Montpellier (UMR 5243), University of Montpellier, 34090 Montpellier, France}

\correspondingauthor{Daniel M. Tartakovsky}{tartakovsky@stanford.edu}

%%%%%%%%%%%%%%%%%%%%% KEYPOINTS %%%%%%%%%%%%%%%%%%%%%%%%%%%
\begin{keypoints}
\item We present a Bayesian inference strategy to estimate Discrete Fracture Network properties from thermal experiments.
\item A neural network surrogate is used to accelerate simulations of heat tracer migration, facilitating exploration of the parameter space.
%\item Adding non-uniform prior information of the DFN parameters can further narrow down the posterior distribution of in uncertainty quantification. DR
\item Prior knowledge about DFN properties sharpens their estimation, yielding a parameter-space region wherein they lie with high probability.
\end{keypoints}
%%%%%%%%%%%%%%%%%%%%%% ABSTRACT %%%%%%%%%%%%%%%%%%%%%%%%%%%
\begin{abstract}
Field-scale properties of fractured rocks play crucial role in many subsurface applications, yet methodologies for identification of the statistical parameters of a discrete fracture network (DFN) are scarce. We present an inversion technique to infer two such parameters, fracture density and fractal dimension, from cross-borehole thermal experiments data. It is based on a particle-based heat-transfer model, whose evaluation is accelerated with a deep neural network (DNN) surrogate that is integrated into a grid search. The DNN is trained on a small number of heat-transfer model runs, and predicts the cumulative density function of the thermal field. The latter is used to compute fine posterior distributions of the (to-be-estimated) parameters. Our synthetic experiments reveal that fracture density is well constrained by data, while fractal dimension is harder to determine. Adding non-uniform prior information related to the DFN connectivity improves the inference of this parameter.
\end{abstract}

% \begin{keyword}
% %% keywords here, in the form: keyword \sep keyword
% Parameter estimation \sep fracture network \sep DFN \sep inverse \sep neural network  
% %% MSC codes here, in the form: \MSC code \sep code
% %% or \MSC[2008] code \sep code (2000 is the default)
% \end{keyword}

% \end{frontmatter}

% \linenumbers

%%%%%%%%%%%%%%%%%%%%%% I. INTRODUCTION %%%%%%%%%%%%%%%%%%%%%%%%%%%
\section{Introduction}\label{intro}

Characterization of fractured rock is a critical challenge in a wide variety of research fields and applications, such as extraction, management and protection of water  resources. In fractured-rock aquifers, fractures can act as preferential flow paths that increase the risk of rapid contaminant migration over large distances. While the resource is generally stored in the surrounding matrix, fractures often determine the spatial extent of the extraction area (the cone of depression or well capture zone). Similar considerations play an important role in (oil/gas and geothermal) reservoir engineering, carbon sequestration, etc.

Various characterization techniques provide complementary information about fractured rocks. These typically rely on direct observation data, surface and borehole data acquired with geophysical techniques, and borehole data collected during hydraulic and tracer experiments \cite{Bonnet2001, dorn2012inferring, Dorn2013, demirel2018characterizing, Roubinet2018}. We focus on the latter because they provide information that is directly related to the hydrogeological structures that drive flow and transport processes. For example, measurements of vertical flow velocities in a borehole under ambient and forced hydraulic conditions are used to estimate the properties of individual fractures that intersect the borehole \cite{Klepikova2013, paillet2012cross, roubinet2015development}, and piezometric data collected in observation boreholes allow one to evaluate features of complex fracture configurations \cite{fischer2018hydraulic, le2010inverse, lods2020groundwater}.  Chemical tracer experiments, typically comprising the interpretation of breakthrough curves, yield information on the short and long paths in the fractured rock; these characterize the discrete fracture network (DFN) and matrix block properties, respectively \cite{roubinet2013particle, haddad2014application}. 

Heat has also been utilized to identify the presence of fractures intersecting boreholes \cite{pehme2013enhanced, read2013characterizing}, to estimate their properties \cite{klepikova2014passive}, and to study flow channeling and fracture-matrix exchange at the fracture scale \cite{de2018thermal, klepikova2016heat}. Most of these thermal experiments employ advanced equipment, which deploys the active line source (ALS) to uniformly modify water temperature in a borehole \cite{pehme2007active} and the distributed temperature sensing (DTS) to simultaneously monitor the resulting temperature changes in observation boreholes   \cite{read2013characterizing}. Thermal tracer experiments offer several advantages over their chemical counterparts.  They do rely on neither localized multi-level sampling techniques nor localized tracer injection in boreholes; they interrogate larger area because heat conduction covers larger area than solute diffusion; and they are not restricted by environmental constraints whereas chemical tracers may remain in the environment for a long time \cite{Akoachere2011, ptak2004tracer}. 

Without exception, the interpretation of hydraulic and tracer experiments involves inverse modeling. The choice of a strategy for the latter depends on the properties of interest, the data considered, the models available to reproduce the data, and the prior information about the studied environment. For canonical fracture configurations between two boreholes, (semi-)analytical and numerical models can be used to the cross-borehole flow-meter experiments mentioned above to evaluate the transmissivity and storativity of the fractures that intersect the boreholes at known depths \cite{Klepikova2013, paillet2012cross, roubinet2015development}; the inversion consists of the gradient-based minimization of a discrepancy between the model's predictions and the collected data. Large-scale systems with complex fracture configurations require the use of sophisticated inversion strategies designed for high volumes of data. Most of such studies generate data via hydraulic and/or tracer tomography experiments, and use the inversion to identify the geometrical and hydraulic properties of  a fracture network \cite{fischer2018hydraulic, le2010inverse, somogyvari2017synthetic}. Very few studies attempt to infer the statistical characteristics of a network, such as fracture density and scaling exponents in distributions of length, orientation and aperture \cite{jang2008inverse, jang2013oil}.

Yet, such statistics are necessary to quantify uncertainty in predictions of hydraulic and transport processes in fractured rocks. Their identification rests on ensem\-ble-based computation, which involves repeated solves of a forward model. Two complementary strategies for making the inversion feasible for large, complex problems are i) to reduce the number of forward solves that are necessary for the inversion algorithm to converge, and ii) to reduce the computational cost of an individual forward solve. The former strategy includes the development of accelerated Markov chain samplers, Hamiltonian Monte Carlo sampling, iterative local updating ensemble smoother, ensemble Kalman filters, and learning on statistical manifolds \cite{barajas2019efficient, boso-2020-learning, boso-2020-data, kang2021hydrogeophysical, zhou2021markov}. The latter strategy aims to replace an expensive forward model with its cheap surrogate/emulator/reduced-order model \cite{ciriello-2019-distribution, lu-2020-lagrangian, lu-2020-prediction}. Among these techniques, various flavors of deep neural %network
networks (DNNs) have attracted attention, in part, because they %remaining
remain robust for large numbers of inputs and outputs \cite{zhou2021markov, mo2020integration, kang2021hydrogeophysical}. Another benefit of DNNs is that their implementation in open-source software is portable to advanced computer architectures, such as graphics processing units and tensor processing units, without significant coding effort from the user.

We combine these two strategies for ensemble-based computation to develop an inversion method, which makes it possible to infer the statistical properties of a fracture network from cross-borehole thermal experiments (CBTEs). 
%We want to take advantage of the recent advances in inversion strategies applied to environmental issues in order to analyze synthetic data corresponding to cross-borehole thermal experiments (CBTEs). Our objective is to evaluate the information contained in CBTE regarding large-scale statistical properties that are used to characterize and generate discrete fracture networks. 
To alleviate the high cost of a forward model of hydro-thermal experiments, we use a meshless, 
particle-based method to solve the two-dimensional governing equations for fluid flow and heat transfer in DFNs (Section~\ref{sec:forward_model}). These solutions, obtained for several realizations of the DFN parameters, are used in Section~\ref{sec:nnm} to train a DNN-based surrogate. The latter's cost is so negligible as to enable us to deploy a fully Bayesian inversion (Section~\ref{sec:inv_strat}) that, unlike ensemble Kalman filter, does not require our quantity of interest to be (approximately) Gaussian. Our numerical experiments, reported in 
%
%To do so, we generate two-dimensional synthetic fractal fracture network models in which hydro-thermal experiments are simulated with fluid flow and heat transfer physically-based models. The computational burden of the forward models is overcome by determining neural network surrogate models that are used for inversion analysis. The inverse problem aiming at evaluating the fracture density and fractal dimension of the fracture networks from this data, is conducted with Bayesian inference, which is well suited for low dimensional problems. The structural and physically-based forward models are presented in Section~\ref{sec:forward_model}, the neural network model formulation in Section~\ref{sec:nnm} and the inversion strategy in Section~\ref{sec:inv_strat}. Numerical experiments are conducted in 
Section~\ref{sec:nume_results}, show that our approach is four orders of magnitude faster than the equivalent inversion based on the physics-based model. These synthetic experiments also reveal that the CBTE data allow one to obtain accurate estimates of fracture density, while the inference of a DFN's fractal dimension is less robust. Main conclusions of this study are summarized in Section~\ref{sec:conclusion}, together with a discussion of alternative strategies to improve the estimation of fractal dimension.

%%%%%%%%%%%%%%%%%%%%%% II. FORWARD MODELS %%%%%%%%%%%%%%%%%%%%%%%%%%%
\section{Models of fracture networks and transport phenomena}
\label{sec:forward_model}

A forward model of CBTEs consists of a fracture network model and those of fluid flow and heat transfer. These models are described in Sections~\ref{sec:fracture_network}, \ref{sec:flow}, and~\ref{sec:heat}, respectively.

%%%%%%%%%%%%%%%%%%%%%% II.1. FRACTURE NETWORK GENERATION %%%%%%%%%%%%%%%%%%%%%%%%%%%

\subsection{Model of fracture networks}
\label{sec:fracture_network}

To be specific, we conceptualize a DFN via the fractal model of \cite{watanabe1995fractal},
\begin{equation}\label{eq:nb_fract}
N_r = C r^{-D},
\end{equation}
that postulates a power-law relationship between the number of fractures, $N_r$, and their relative length $r$ (normalized by smallest fracture length $r_0$), in a domain of characteristic length $L$. The parameters $C$ and $D$ denote fracture density and fractal dimension, respectively. If a network's smallest fracture has length $r_0$, then the number of classes in the WT model is $N_f = \mbox{int}(C/r_0^D)$ and the relative length of fractures in the $i$th class is $r_i = (C/i)^{1/D}$ ($i=1,\dots,N_f$). This formulation is equivalent to the model \cite{davy1990some} that expresses fracture density $n(l,L) = \alpha L^{\mathcal D} l^{-a}$ in terms of fracture length $l$ and domain size $L$, if one sets $\alpha = C D/N_f$, $\mathcal D=D$, and $a=D+1$. The latter model reproduces self-similar structures observed in numerous studies \cite[chapter 6.6.8]{Sahimi2011}, allowing one to represent realistic fracture networks with the minimal number of parameters. 

To generate a synthetic data set, we consider fractures arranged at two preferred angles $\theta_1 = 25^\circ$ and $\theta_2 =145^\circ$ in a $100\times 100$~m$^2$ domain. Fracture centers are randomly distributed over the whole domain, and their aperture is set to $5 \times 10^{-4}$~m, as in \cite{gisladottir2016particle}. The resulting DFN is simplified by removing the fractures that are not, directly or indirectly through other fractures, connected to the domain's perimeter. Fluid flow and heat transfer are modeled on this fracture network backbone.

%%%%%%%%%%%%%%%%%%%%%% II.2. FLUID FLOW MODEL %%%%%%%%%%%%%%%%%%%%%%%%%%%
\subsection{Model of fluid flow in fracture networks}
\label{sec:flow}

We deploy a standard model of single-phase steady-state laminar flow in a DFN, which assumes the ambient rock matrix to be impervious to fluid. The flow of an incompressible fluid is driven by a hydraulic head gradient, $J$, due to constant hydraulic heads imposed on the left and right boundaries, the top and bottom boundaries are impermeable. 

The fracture extremities and intersections of the DFN, whose construction is detailed above, form the network nodes and a fracture connecting two adjacent nodes are referred to as the network edge. Flow rate in each edge is computed as the cross-sectional average of the Poiseuille velocity profile. Thus, the flow rate, $u_{ij}$, of flow from node $i$ to node 
$j$ is $u_{ij} = - b_{ij}^2/(12 \nu) J_{ij}$, where $\nu$ is the fluid's kinematic viscosity, $b_{ij}$ is the aperture of the fracture connecting the nodes $i$ and $j$, and 
$J_{ij} = (h_j - h_i)/l_{ij}$ is the hydraulic head gradient between these nodes with $l_{ij}$ the distance between these nodes. The hydraulic heads at the DFN nodes, $h_i$ ($i=1,2,\dots$), are computed as the solution of a linear system built by enforcing mass conservation at each node: $\sum_{k \in \mathcal N_i} b_{ki} u_{ki}=0$, where $\mathcal N_i$ is the set of the neighboring nodes of node $i$ (see, e.g., \cite{gisladottir2016particle, zimmerman-2020-solute} for details).

%%%%%%%%%%%%%%%%%%%%%% II.3. HEAT TRANSFER MODEL %%%%%%%%%%%%%%%%%%%%%%%%%%%
\subsection{Model of heat transfer in fractured rock}
\label{sec:heat}

The DFN backbone constructed in Section~\ref{sec:fracture_network} is further pruned by removing the edges representing the fractures with negligible flow velocities, e.g., $u_{ij} \le 10^{-10}$~m/s used in the subsequent numerical experiments.  %This leads to removing the "dead-ends" of the fracture networks, which correspond to fracture segments for which one of the extremities has only one neighboring node. 
Convection in the resulting fracture network and conduction in the host matrix rock are modeled via the particle-based approach~\cite{gisladottir2016particle}. The computational cost of this method is significantly lower than that of its mesh-based alternatives because it discretizes only the fracture segments, while the matrix is not meshed. The particle displacement is associated with conduction and convection times in the fracture and the matrix, respectively. The latter time is defined from analytical solutions to a transport equation for fracture-matrix systems, and truncated according to the probability $p_\text{lim}$ for the particle to reach a neighboring fracture by conduction through the matrix. %The fracture segment discretization is defined through the parameter $p_\text{lim}$ that denotes this probability. 
Complete mixing is assumed at the fracture intersections, implying that the probability for a particle to enter into a fracture depends only on the flow rate arriving at the considered node. 

CBTEs are simulated by uniformly injecting $N_\text{part}$ particles on the left side of the domain and recording their arrival times on the right side. The cumulative distribution functions (CDFs) of these arrival times describe the changes in the relative temperature $T^*$ observed at distance $L$ from the heat source, assuming complete mixing in the vertical direction at the observation position. The relative temperature is defined as $T^*=(T_\text{obs}-T_\text{in})/(T_\text{inj}-T_\text{in})$, where $T_\text{in}$ is the initial (at $t=0$) fluid temperature in the system, and $T_\text{inj}$ and $T_\text{obs}$ the temperature at the injection and observation positions, respectively \cite{gisladottir2016particle}.

%%%%%%%%%%%%%%%%%%%%%% PROBLEM DEFINITION %%%%%%%%%%%%%%%%%%%%%%%%%%%
\section{Neural network model formulation}
\label{sec:nnm}

%\subsection{Forward surrogate model}

We define a NN surrogate for the physics-based model described in Section~\ref{sec:forward_model} with a map, 
\begin{linenomath*}
\begin{equation}\label{eq:forw_surr1}
    \mathbf{f}  : (C, D) \rightarrow F(x), \quad F(x) = \mathbb{P}(X\le x), \quad x \in \mathbb{R},
\end{equation}
\end{linenomath*}
where $(C, D)$ are the fracture network parameters, and $F(x)$ is the CDF of a particle's arrival time $X$, i.e., the probability that $X$ does not exceed a certain value $x$. Since the nonzero probability space of $F(x)$ varies for different simulations %\cite[e.g.,][and Section~\ref{sec:nume_results} below]{gisladottir2016particle,Ruiz2014}, DR
\cite[and Section~\ref{sec:nume_results} below]{gisladottir2016particle, Ruiz2014}, 
we find it convenient to work with the inverse CDF (iCDF) $F^{-1}$. Because any CDF is a continuous monotonically increasing function, the iCDF (or quantile CDF) is defined as
\begin{linenomath*}
\begin{equation}\label{eq:forw_surr2}
\text{iCDF}: Q(p)=F^{-1}(p) = \min\{x\in \mathbb{R}: F(x)\ge p \}, \quad p\in (0,1).
\end{equation}
\end{linenomath*}
If $Q(p)$ is discretized into a set of $N_k$ quantiles $\{p_1, \dots, p_{N_k} : 0<p_1< \dots < p_{N_k} <1 \}$, then
\begin{linenomath*}
\begin{align}\label{eq:forw_surr3}
\begin{split}
 \text{iCDF} = \{ Q(p_1), \dots, Q(p_{N_k}) \}, \qquad
 Q(p_1) < \dots <Q(p_{N_k}).
\end{split}
\end{align}
\end{linenomath*}

%\subsection{Fully connected NNs}
%\label{sec:fc_nn}

Consider a fully connected neural network (FCNN) 
\begin{linenomath*}
\begin{equation}\label{eq:FCNN}
\mathbf{NN:m}\xrightarrow{\text{FCNN}} \mathbf{\hat d}
\end{equation}
\end{linenomath*}
that describes the forward surrogate model~\eqref{eq:forw_surr1}--\eqref{eq:forw_surr3}. The vector $\mathbf{m}$, of length $N_m$, contains the parameters to be estimated (in our problems, these parameters are $C$ and $D$, so that $N_m=2$); and the vector $\mathbf{\hat d}$, of length $N_d$, contains the discretized values of the iCDF computed with the model $\mathbf{NN}$. This model is built by defining an $N_d \times N_m$ matrix of weights $\mathbf W$, whose values are obtained by minimizing the discrepancy between the vectors ${\mathbf{\hat d}}$ and the vector $\mathbf d$ comprising the output of physics-based model from Section~\ref{sec:forward_model}. Since the relationship between $\mathbf{m}$ and $\mathbf{d}$ is likely to be highly nonlinear, we relate $\mathbf{m}$ and ${\mathbf{\hat d}}$ via a nonlinear model $\hat{\mathbf{d}} = \sigma(\mathbf W \mathbf m)$, in which the prescribed ``activation'' function $\sigma(\cdot)$ operates on each element of the vector $\mathbf W \mathbf m$. Commonly used activation functions include sigmoid functions (e.g., $\tanh$) and the rectified linear unit (ReLU). The latter, $\sigma(s) = \max(0, s)$, is used in this study due to its proven performance in similar applications \cite{agarap2018deep, zhou2021markov, mo2019deep}. 

The nonlinear regression model $\hat{\mathbf{d}} = \sigma(\mathbf W \mathbf m) \equiv (\sigma \circ \mathbf{W})(\mathbf m)$ constitutes a single layer in a NN. A (deep) FCNN model with $N_l$ layers is constructed by a repeated application of the activation function to the input,
\begin{subequations}
\begin{linenomath*}\label{eqa:fc_net}
\begin{equation}
\hat{\mathbf{d}} = \mathbf{NN}(\mathbf m; \boldsymbol\Theta) \equiv (\sigma_{N_l} \circ \mathbf{W}_{N_l - 1})\circ \ldots \circ (\sigma_{2} \circ \mathbf{W}_{1})(\mathbf m).
\end{equation}
\end{linenomath*}
%
%Practically, various activation functions might be used in one neural network. 
The parameter set $\boldsymbol\Theta = \{ \mathbf W_1,\ldots,\mathbf W_{N_l-1} \}$ consists of the weights $\mathbf W_n$ connecting the $n$th  and $(n+1)$st layers with the recursive relationships
\begin{linenomath*}
\begin{align}
\begin{cases}
    \mathbf{s}_1 = (\sigma_{2} \circ \mathbf{W}_{1})(\mathbf m)  \equiv \sigma_2(\mathbf{W}_{1} \mathbf m), \\
    \mathbf{s}_i = (\sigma_{i+1} \circ \mathbf{W}_{i})(\mathbf{s}_{i-1})  \equiv \sigma_{i+1}(\mathbf{W}_{i} \mathbf{s}_{i-1}), \quad i=2,\dots,N_l-2\\
    \hat{\mathbf{d}} = (\sigma_{N_l} \circ \mathbf{W}_{N_l-1})(\mathbf{s}_{N_l - 2}) \equiv \sigma_{N_l}(\mathbf{W}_{N_l -1} \mathbf{s}_{N_l - 2}).
\end{cases}
\label{eqa:fc}
\end{align}
\end{linenomath*}
\end{subequations}
Here, $\mathbf s_i$ is the vector of data estimated in the $i$th layer; $\mathbf W_1$, $\mathbf W_i$  ($i=2,\dots,N_l-2$) and $\mathbf{W}_{N_l-1}$ are the matrices of size $d_1 \times N_m$, $d_i \times d_{i-1}$ and $N_d \times d_{N_l-2}$, respectively; and the integers $d_i$ ($i=1,\dots,N_l-2$) represent the number of neurons in the corresponding inner layers of the NN. 
%Sometime, bias parameters $\mathbf{b_i}$ with dimension $d_i\times d_{i-1}$ is added to $\mathbf{s_i}$ too, these $\mathbf{b}$ are included into $\mathcal \Theta$ as fitting parameters.
%
The fitting parameters $\boldsymbol\Theta$ are obtained by minimizing the discrepancy (or ``loss function'') $\mathcal L(\mathbf{d}_i, \mathbf{\hat{d}}_i)$ between $\mathbf{\hat{d}}$ and $\mathbf{d}$,
\begin{linenomath*}
\begin{equation}\label{eqa:mini_loss}
    \boldsymbol\Theta = \underset{\boldsymbol\Theta}{\mathrm{argmin}} \sum_{i=1}^{N_{\text{data}}} \mathcal L(\mathbf{d}_i, \mathbf{\hat{d}}_i), \qquad \hat{\mathbf{d}_i} = \mathbf{NN}(\mathbf{m}_i; \boldsymbol\Theta), 
\end{equation}
\end{linenomath*}
where $N_\text{data}$ is the number of forward runs of the physics-based model. We use the stochastic gradient descent optimizer \cite{ruder2016overview} to carry out this step, which is commonly referred to as ``network training''.

%\subsection{Neural network model setting}
A choice of the functional form of the loss function $\mathcal L$ affects a NN's performance. %Measuring the goodness of the fitting for the fracture forward model requires a good choice of the loss function $L$ used in Equation \eqref{eqa:mini_loss}. This function computes the divergence between the predicted and computed TICDFs. 
Studies on measuring quantile divergence, especially for discrete inverse distribution, are scarce. Measures of the distance between probability distributions, such as the Kullback-Leibler (KL) divergence \cite{kullback1997information} and the Hellinger distance \cite{le2012asymptotic}, might or might not be appropriate for inverse distributions. Thus, while the KL divergence is a popular metric in Bayesian inference \cite{boso-2020-learning} and generative NNs \cite{kingma2013auto, goodfellow2014generative}, its asymmetry precludes its use in~\eqref{eqa:mini_loss}. Consequently, we quantify the distance between two discrete distributions $P=(p_1, \dots, p_{N_k})$ and $P' = (p'_1, \dots, p'_{N_k})$ in terms of the Hellinger distance,
\begin{linenomath*}
\begin{align}
\begin{split}
\mathcal L_\text{H}(P, P') = \frac{1}{\sqrt{2}} \|\sqrt{P} - \sqrt{P'} \|_{2}
= \left( \frac{1}{2} \sum^{N_k}_{i=1}(\sqrt{p_i} - \sqrt{p'_i})^2 \right)^{1/2},
\end{split}
\end{align}
\end{linenomath*}
i.e., solve the minimization problem~\eqref{eqa:mini_loss} with $\mathcal L \equiv \mathcal L_\text{H}(Q,\hat Q)$.

To reduce the training cost and improve the NN's performance, we specify additional features to refine the initial guess of input parameters. The relationships between the fractal DFN parameters in Section~\ref{sec:fracture_network}, suggest the choice of $C^{1/D}$, $C^{-D}$ and $CD$ (which are equal to $r_i i^{1/D}$, $r_0 / N_f^D$ and $\alpha N_f$, respectively) and $1/D$ as extra input features. Given the pair of initial parameters $(C,D)$, the resulting full set of  parameters for the NN is
\begin{linenomath*}
\begin{equation}\label{eq:mNN}
\mathbf{m}_\text{NN} = (C, D, C^{1/D}, C^{-D}, CD, 1/D)^\top.
\end{equation}
\end{linenomath*} 

%%%%%%%%%%%%%%%%%%%%%%%%%%%%%%%%%%%%%%%%%%%%%%% INVERSION STRATEGY %%%%%%%%%%%%%%%%%%%%%%%%%%%%%%%%%%%%%%%%%%%%%%%%%%%%%%%%%%%%%%%%%%
\section{Inversion via Bayesian update}
\label{sec:inv_strat}

%\subsection{Theoretical background}
%Considering Bayesian inference, the posterior distribution $f_{\mathbf m | \mathbf d}$ of the parameter vector $\mathbf{m}$, which is the refined distribution given the data $\mathbf{d}$, is defined as
According to the Bayes rule, the posterior probability density function (PDF) $f_{\mathbf m | \mathbf d}$ of the parameter vector $\mathbf{m}$ is computed as
\begin{linenomath*}
\begin{equation}\label{eqa:post}
f_{\mathbf m | \mathbf d} (\tilde{\mathbf m}; \tilde{\mathbf d}) = \frac{f_\mathbf{m}(\tilde{\mathbf m}) f_{\mathbf d | \mathbf m}(\tilde{\mathbf m}; \tilde{\mathbf d})}{f_\mathbf{d}(\tilde{\mathbf d})}, \qquad f_\mathbf{d}(\tilde{\mathbf d}) = \int f_\mathbf{m}(\tilde{\mathbf m}) f_{\mathbf d | \mathbf m}(\tilde{\mathbf m}; \tilde{\mathbf d}) \text d \tilde{\mathbf m},
\end{equation}
\end{linenomath*}
where $\tilde{\mathbf d}$ and $\tilde{\mathbf m}$ are the deterministic coordinates of random variable $\mathbf d$ and $\mathbf m$, respectively; $f_{\mathbf{m}}$ is the prior PDF of $\mathbf{m}$;  $f_\mathbf{d|m}$ is the likelihood function (i.e., the joint PDF of the measurements conditioned on the model predictions, which is treated as a function of $\mathbf{m}$); and the normalizing factor $f_{\mathbf{d}}$ ensures that $f_{\mathbf m | \mathbf d}$ integrates to 1.

We take the likelihood function $f_\mathbf{d|m}$ to be Gaussian, 
\begin{linenomath*}
\begin{equation}\label{eqa:likelihood}
f_{\mathbf d | \mathbf m}(\tilde{\mathbf m}; \tilde{\mathbf d}) = \frac{1}{\sigma_{\text{d}} \sqrt{2\pi} } \exp\left[ -\frac{1}{2} \frac{L_{H}(\mathbf{\tilde{d}, g(\tilde{m}))}}{\sigma_{\text{d}}^2}\right].
\end{equation}
\end{linenomath*}
This PDF has the standard deviation $\sigma_{\text{d}}$ and is centered around the square root of the Hellinger distance between the data $\tilde{\mathbf d}$ predicted by the likelihood and the data $\mathbf{ g(\tilde{m})}$ provided by the forward model $\mathbf g$. Addition of prior knowledge of $\mathbf{m}$ to the likelihood function is done within the standard Bayesian framework by assuming that the prior PDF is as important as the data. We explore how the posterior PDF can be improved by adjusting the impact of the prior. To do so, we treat the latter as a regularization term with a tunable hyper-parameter $\gamma$ that corresponds to the weight associated with the prior, enabling us to reduce the impact of the prior when its knowledge does not seem to be persuasive. The resulting posterior PDF is formulated as
\begin{linenomath*}
\begin{align}\label{eq:gamma}
f_{\mathbf m | \mathbf d} (\tilde{\mathbf m}; \tilde{\mathbf d}) & \propto \text{e}^{ -H(\mathbf{\tilde{m}}) }, \qquad H(\mathbf{\tilde{m}}) = H_{\text{obs}}(\mathbf{\tilde{m}}) + \gamma H_{\text{reg}}(\mathbf{\tilde{m}}),
\end{align}
\end{linenomath*}
where $H_{\text{obs}}(\mathbf{\tilde{m}}) = -\log(f_{\mathbf d | \mathbf m}(\tilde{\mathbf m};\tilde{\mathbf d}))$ and $H_{\text{reg}}(\mathbf{\tilde{m}}) = -\log(f_\mathbf{m}(\tilde{\mathbf m}))$ are the negative log-likelihood and log-prior distributions, respectively. This yields
\begin{linenomath*}
\begin{align}\label{eq:final_prior}
f_{\mathbf m | \mathbf d} (\tilde{\mathbf m}; \tilde{\mathbf d})  \propto f_{\mathbf d | \mathbf m}(\tilde{\mathbf m};\tilde{\mathbf d}) \left( f_\mathbf{m}(\tilde{\mathbf m}) \right)^{\gamma}, \quad \gamma \in [0, 1].
\end{align}
\end{linenomath*}
This posterior PDF is computed via the following algorithm.

%\subsection{Numerical implementation}\label{sec:method_grid}
%The efficiency of the neural network model that is previously defined enables us to implement a robust inverse analysis by computing the posterior distribution of a large number of inputs. The corresponding Monte Carlo algorithm is implemented as follows:
%The posterior distribution previously defined is computed with the following Monte Carlo algorithm:
\begin{enumerate}
\item The domains $\mathcal C$ and $\mathcal D$ of values for the parameters $C$ and $D$ are discretized with $N_C$ and $N_D$ nodes, respectively. The result is a $N_C \times N_D$ regular grid for the parameter pair $(C,D)$ with coordinate vectors $\mathbf m_{ij}=(C_i,D_j)^\top$ ($i=1,\dots, N_C$, $j = 1,\dots,N_D$).

\item The iCDFs~\eqref{eq:forw_surr3} are computed with the forward model $\mathbf g$ for all pairs $\mathbf m_{ij}$.
%\textcolor{blue}{[ZT: compute with the neural network, because $N_C, N_D$ are going to be very large, I computed 1000000 pairs.]}

\item The negative log-likelihood $H_{\text{obs}}(\mathbf{m}) = -\ln(f_{\mathbf d | \mathbf m}(\tilde{\mathbf m};\tilde{\mathbf d}))$ is computed via~\eqref{eqa:likelihood}, with the data $\mathbf g(\textbf m)$ provided by model $\textbf g$ in Step 2.

\item The posterior PDF $f_{\mathbf m | \mathbf d}$ is computed via~\eqref{eq:final_prior} by adjusting the weight $\gamma$ assigned to the prior knowledge. (The case $\gamma=0$ corresponds to a uniform prior for $\mathbf{m}$, where the unnormalized posterior PDF is equivalent to the likelihood.)
\end{enumerate}

This brute-force implementation of Bayesian inference is only made possible by the availability of the FCNN surrogate, whose forward runs carry virtually zero computational cost. In its absence, or if the number of unknown parameters were large, one would have to deploy more advanced Bayesian update schemes such as Markov chain Monte Carlo~\cite{zhou2021markov,barajas2019efficient} or ensemble updating methods~\cite{mo2019deep, mo2020integration}.

%%%%%%%%%%%%%%%%%%%%%% III. RESULTS %%%%%%%%%%%%%%%%%%%%%%%%%%%
\section{Numerical experiments}
\label{sec:nume_results}

The synthetic generation of DFNs and breakthrough times, $t_\text{break}$, for a heat tracer is described section~\ref{sec:experiment}. Generation of the data for CNN training is described in section~\ref{sec:data_generation}, with the construction of a CNN surrogate for the PDE-based model (section~\ref{sec:forward_model}) reported in section~\ref{sec:cnn_training}. In sections~\ref{sec:standard} and~\ref{sec:modified}, we use this surrogate to accelerate the solution of the inverse problem of identifying the DFN properties from the breakthrough-time data.

\subsection{Synthetic heat-tracer experiment}
\label{sec:experiment}

Our synthetic heat tracer experiment consists of injected hot water with temperature $T_\text{inj}$ at the inlet ($x_1 = 0$) and observing temperature changes at the outlet 
($x_1 = L$). The goal is to infer the statistical properties of a DFN, $C$ and $D$, from a resulting breakthrough curve. A fracture network with known values of $C$ and $D$ serves as ground truth, with possible measurement errors neglected. Consistent with~\cite{gisladottir2016particle}, we set the externally imposed hydraulic gradient across the simulation domain to $J = 0.01$ and the thermal diffusion coefficient in the matrix to $D_\text{therm} = 9.16 \times 10^{-7}$~m$^2$/s.

\subsection{Generation and analysis of synthetic data}
\label{sec:data_generation}

To generate data for the CNN training and testing, we considered the WT fracture networks~\eqref{eq:nb_fract} with $C \in [2.5, 6.5]$ and $D \in [1.0, 1.3]$. These parameter ranges are both observed experimentally~\cite{main1990influence, scholz1993fault} and used in previous numerical studies~\cite{gisladottir2016particle, watanabe1995fractal}. The parameter space $[2.5, 6.5] \times [1.0, 1.3]$ was uniformly discretized into $N_\text{sim} = 10^4$ nodes, i.e., pairs of the parameters $(C,D)_i$ with $i=1,\dots,N_\text{sim}$. The number of injected particles, $N_\text{part}$, representing the relative temperature of the injected fluid during a CBTE, $T_\text{inj}$, varied between $10^2$ and $10^4$.

\begin{figure}[htbp]
    \centering
    \includegraphics[width=0.8\textwidth]{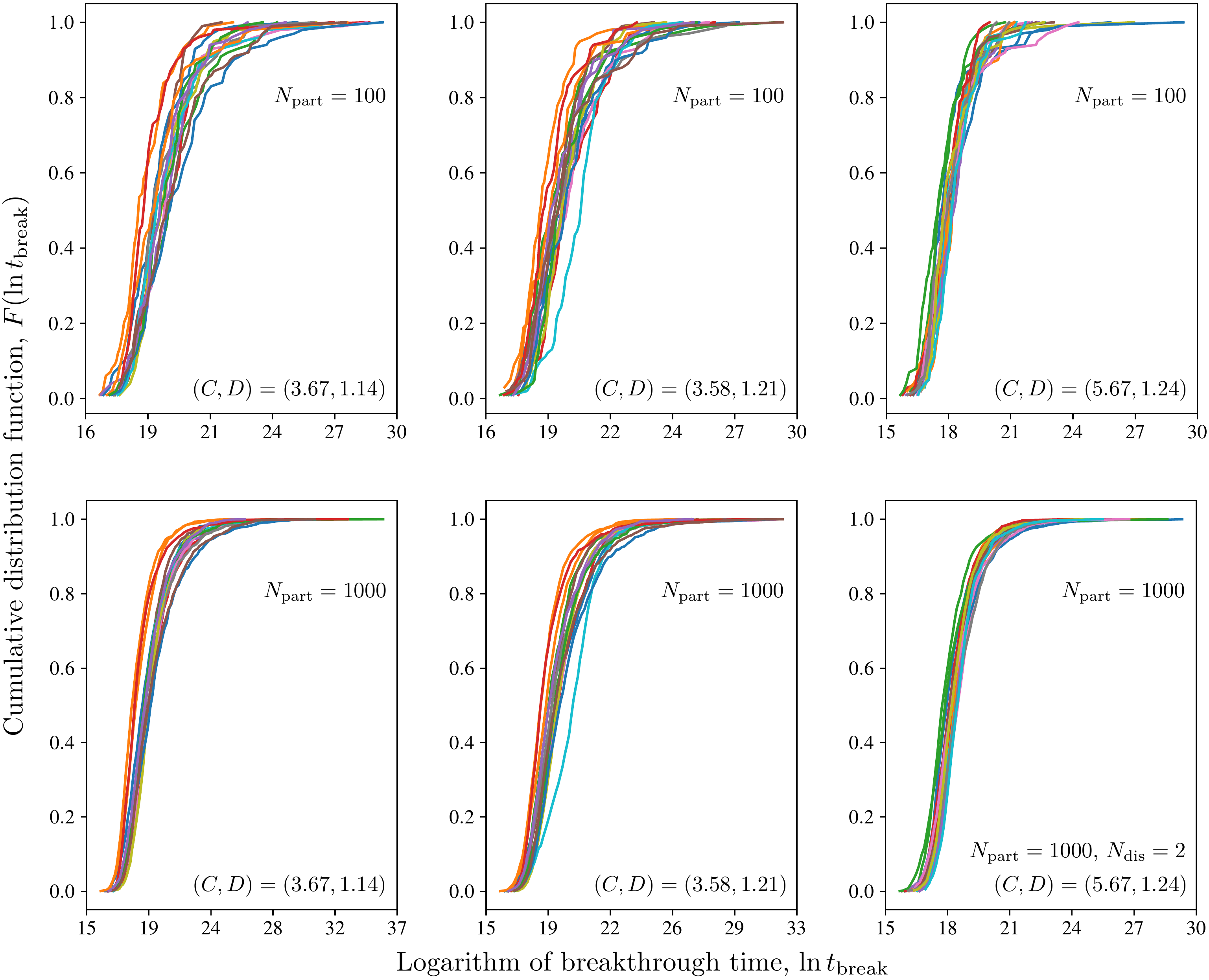}
    \caption{Representative CDFs of the logarithm of breakthrough times (in seconds) of $N_\text{part}$ particles, $F(\ln t_\text{break})$, for 20  realizations of the DFN characterized by a given combination of the DFN parameters $(C,D)$. Each colored curve corresponds to a different random realization; in all simulations, we set $p_\text{lim} = 0.5$. }
    \label{fig:cdf_combined}
\end{figure}

In addition to $N_\text{part}$, the simulation time and accuracy of each forward model run are largely controlled by the number of elements used to discretize a fracture, which is defined by the parameter $p_\text{lim}$ introduced in section~\ref{sec:heat}. The simulation time $t_\text{sim}$ refers to the time (in seconds) it takes to estimate the CDF of breakthrough times for one random DFN realization and one of the $N_\text{sim} = 10^4$ pairs of the parameters $(C,D)$. We found the average $t_\text{sim}$ not to exceed 1~s if either $N_\text{part} = 100$ or the fracture is not discretized (Table 1 %\ref{tab:time} 
 of the Supplemental Material); the average is over 20 random realizations of the DFN  obtained with different random seeds for each parameter pair $(C,D)$.

\begin{figure}[htbp]
    \centering
    \includegraphics[width=0.9\textwidth]{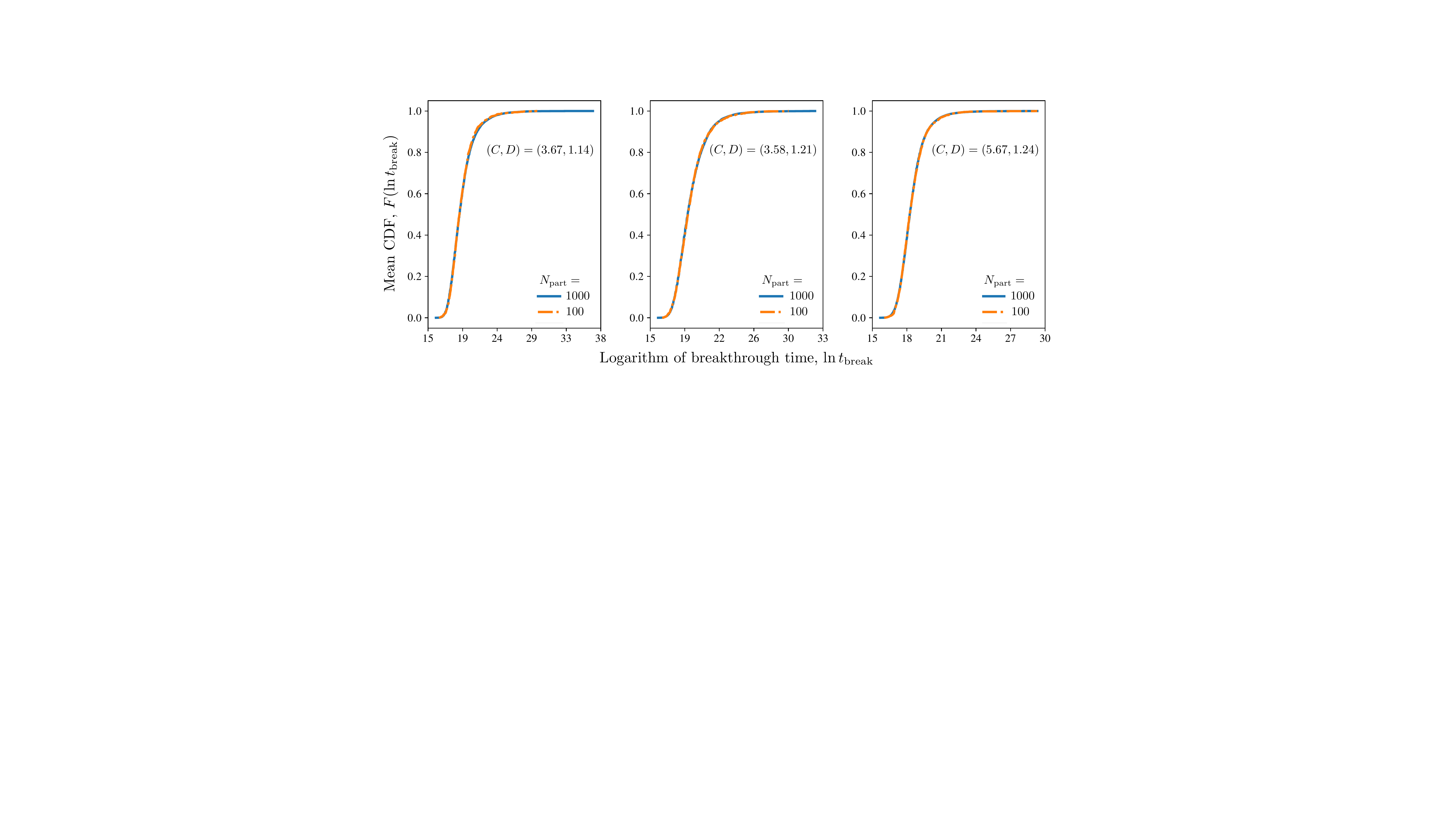}
    \caption{Mean CDFs of the logarithm of breakthrough times (in seconds) of $N_\text{part}$ particles, $F(\ln t_\text{break})$, averaged over the corresponding DFN realizations in Figure~\ref{fig:cdf_combined}.} 
    \label{fig:cdf_ave}
\end{figure}

Representative CDFs of breakthrough times of $N_\text{part}$ particles, in each of these 20 DFN realizations, are displayed in Figure~\ref{fig:cdf_combined} for three pairs of the DFN parameters $(C,D)$. The across-realization variability of the CDFs is more pronounced for $N_\text{part} = 10^2$ then $10^3$ particles, and visually indistinguishable when going from $N_\text{part} = 10^3$ to $10^4$ particles (not shown here). Likewise, no appreciable differences between the CDFs computed with $p_\text{lim} = 0.5$ and $0.2$ were observed. 
Finally, when the random-seed effects are averaged out, the resulting breakthrough-time CDFs for $N_\text{part} = 10^2$ and $10^3$ are practically identical (Figure~\ref{fig:cdf_ave}).
Based on these findings, in the subsequent simulations, we set $N_\text{part} = 100$ and 
$p_\text{lim} = 0.5$ in order to obtain an optimal balance between the computational time and accuracy.

\begin{comment}
\begin{figure}
    \centering
    \includegraphics[width=0.9\textwidth]{Figures/uncorr20seeds.pdf}
    \caption{Examples of TCDFs of 20 random realizations for 6 parameter pairs $(C,D)$ with the simulation parameter set \textit{Simu~7}. Each colored curve corresponds to a different random realization.}
    \label{fig:TCDF_simu7}
\end{figure}

\begin{figure}
    \centering
    \includegraphics[width=0.9\textwidth]{Figures/uncorr20seeds_1000.pdf}
    \caption{Examples of TCDFs of 20 random realizations for 6 parameter pairs $(C,D)$ with the simulation parameter set \textit{Simu 8}. Each colored curve corresponds to a different random realization.}
    \label{fig:TCDF_simu8}
\end{figure}

\begin{figure}[htbp]
    \centering
    \includegraphics[width=0.7\textwidth]{Figures/uncorr_simu_hist.pdf}
    \caption{Histogram of connected and successfully simulated realizations over 20 realizations.}
    \label{fig:hist_uncorr}
\end{figure}
\end{comment}

\begin{figure}[htbp]
    \centering
    \includegraphics[width=0.7\textwidth]{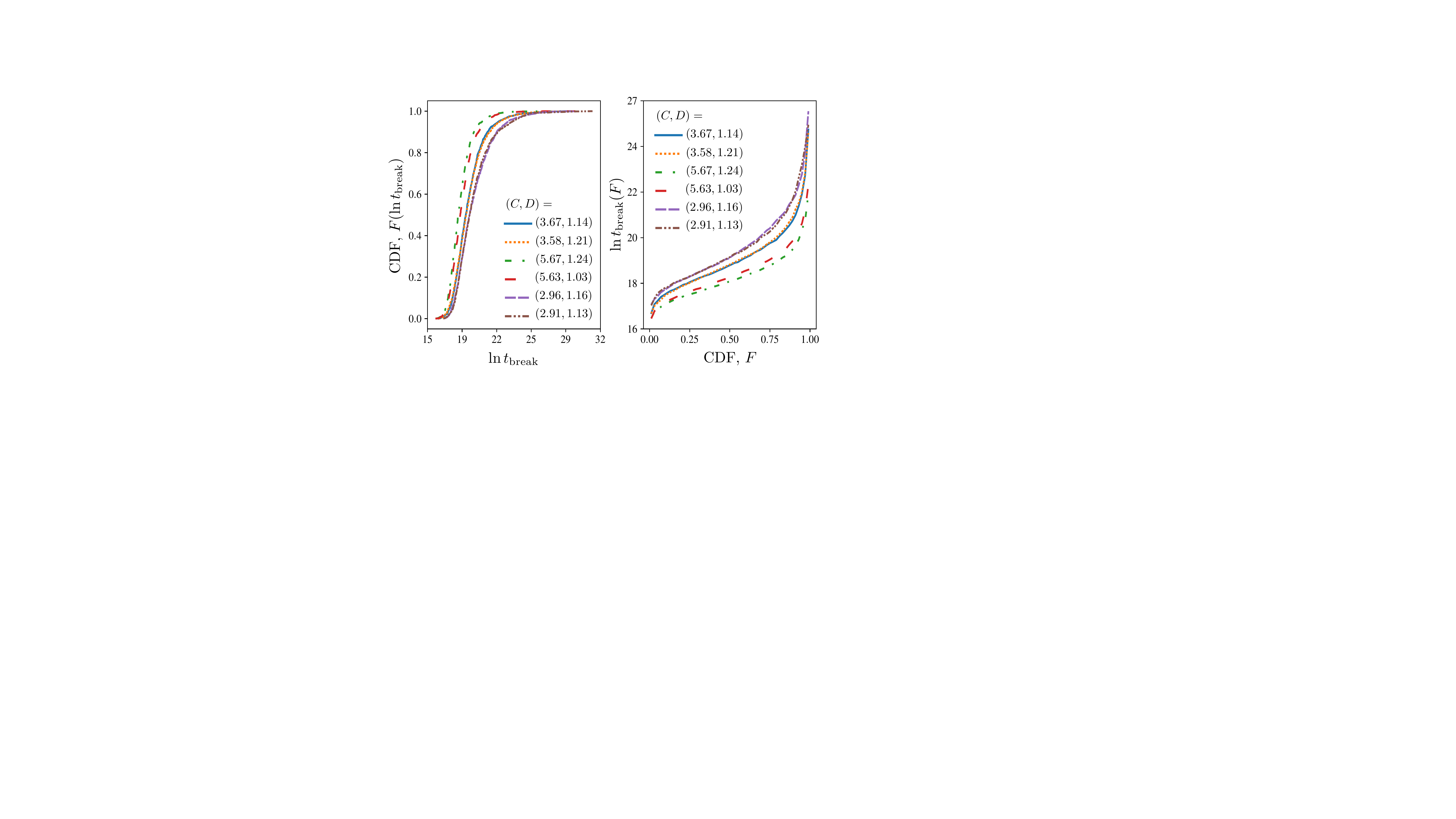}
    \caption{ CDFs (left) and corresponding iCDFs (right) of the thermal breakthrough times for a single realization of the six DFNs characterized by six pairs of the parameters $(C,D)$.}
    \label{fig:CDF_ICDF}
\end{figure}

For some parameter pairs $(C,D)$, not every DFN realization (defined by the random seed) hydraulically connects the injection and observation boundaries. Such hydraulically disconnected networks are not suitable for our flow model (see section~\ref{sec:flow}). %the number of displayed CDFs is smaller than 20 because the physically-based transport model could not perform for all the 20 random fracture networks. This is usually due to the presence of not connected fracture networks for which the fluid flow distribution is not defined. 
However, in our numerical experiments, there were at least 10---and, in the majority of cases, 19---connected fracture networks for each $(C,D)$ pair (Figure~2 %\ref{fig:hist_uncorr}
of the Supplemental Material).

The final step in our data generation procedure consists of converting the estimated CDFs $F$ into corresponding iCDFs $F^{-1}$ (Figure~\ref{fig:CDF_ICDF}). The latter form the data set $\mathbf d$, different parts of which are used to train a CNN and to verify its performance.

\subsection{CNN training and testing}
\label{sec:cnn_training}

The data generated above are arranged in a set $\{\mathbf m_{ \text{NN}_i}, \mathbf{d}_i\}_{i=1}^{N_\text{sim}}$ with $N_\text{sim} = 10^4$ and $\mathbf m_\text{NN}$ defined in~\eqref{eq:mNN}. We randomly select $8 \cdot 10^3$ of these pairs to train the FCNN $\mathbf{NN}$ in~\eqref{eq:FCNN}, leaving the remaining $2 \cdot 10^3$ for testing. The output data $\mathbf d$ come in the form of iCDFs, i.e., non-decreasing series of numbers. Since a NN model is not guaranteed to reproduce this trend, we use the hyper-parameter tuning method~\cite{liaw2018tune} to perform the search in the hyper-parameter space specified in Table~\ref{tab:hyper_search}.

\begin{table}[htbp]
    \caption{Hyper-parameter search space defined by the number of layers, the number of neurons in each layer, the optimizer names, and (logarithm of) the learning rate. These parameters are uniformly sampled from either a discrete set of values, $\mathcal{U}\{\cdot,\cdot,\dots,\cdot\}$, or an interval, $\mathcal{U}[\cdot,\cdot]$. The RMSprop optimizer \cite{graves2013generating,hinton2012neural}, \texttt{rms}; the stochastic gradient descent optimizer \cite{sutskever2013importance}, \texttt{sgd}; the Adagrad optimizer \cite{ duchi2011adaptive}~\texttt{ada}; and the Adam optimizer \cite{kingma2014adam}, \texttt{adam}, slightly differ from each other when performing the parameter gradient descent during the NN training.  }
    \label{tab:hyper_search}
    \centering
    \begin{tabular}{ll}
        \hline
        Parameter name & Search region \\
        \hline
        Number of layers & $\mathcal{U}\{3,4,5,6\}$ \\
        Number of neurons & $\mathcal{U}\{2^2, 2^3,...,2^9\}$ \\
        Optimizer name & $\mathcal{U}\{\texttt{rms}, \texttt{sgd}, \texttt{ada}, \texttt{adam} \}$  \\ 
        Learning rate, $l_r$ & $\log_{10}(l_r) \sim \mathcal{U}[-4,-2]$ \\
        \hline
    \end{tabular}
\end{table}

\begin{figure}[htbp]
    \centering
    \includegraphics[width=0.9\textwidth]{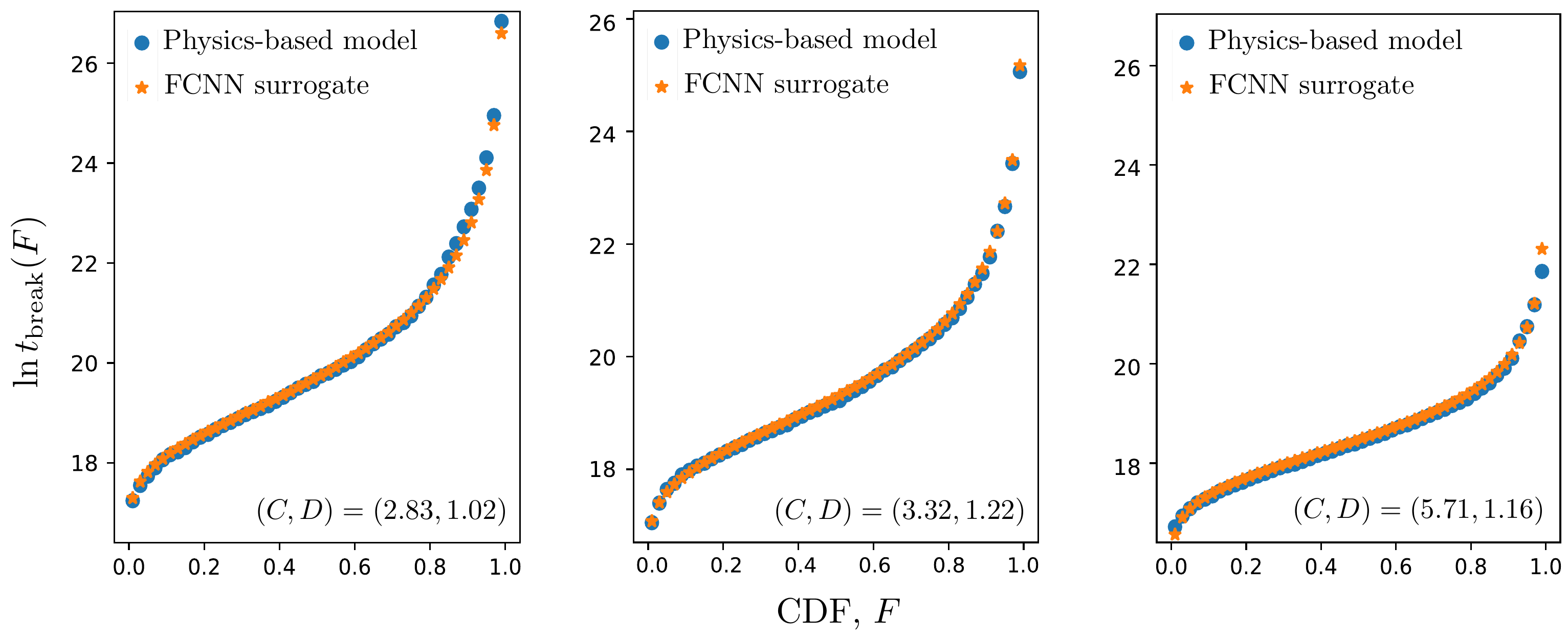}
    \caption{Physics-based and FCNN predictions of the iCDFs of the particle breakthrough times in DFNs characterized by different parameter-pairs $(C,D)$ not used for training.}
    \label{fig:FCNN_test}
\end{figure}

The hyper-parameter search involved 2500 trials; in each trial, the subset of data $\{\mathbf m_{ \text{NN}_i}, $ $\mathbf{d}_i\}_{i=1}^{8000}$ were randomly split into a training set consisting of 6400 pairs $\{\mathbf m_{ \text{NN}_i}, \mathbf{d}_i\}$ and a validation set comprising the remaining 1600 pairs $\{\mathbf m_{ \text{NN}_i}, \mathbf{d}_i\}$. For each epoch, the 6400 training pairs were used to optimize the NN parameters, and the NN accuracy is evaluated on the validation set. Each trial used one of the optimizers in Table~\ref{tab:hyper_search} for at most $10^3$ epochs; the trial was stopped if the validation loss did not decrease for $10^2$ epochs. After completion of all the trials with these rules, the trial with the smallest validation loss was saved. The optimal FCNN, described in Table~\ref{tab:NN_arch}, has 6 layers between the input and output layers and is obtained using the Adam optimizer with the Adam optimizer coefficients $\beta = (0.9, 0.999)$ to perform gradient descent. This trial is associated with a learning rate $l_r = 0.00403$ and the averaged Hellinger loss of $0.0827$ on the validation set. This FCNN was further trained with a learning rate that reduces on plateau of the validation performance to further fine-tune the model parameters for another $10^3$ epochs; the ending testing Hellinger loss is $0.0652$ and the total training time is $37340$ seconds. Figure~\ref{fig:FCNN_test} depicts the FCNN predictions of the iCDFs of the particle breakthrough times in DFNs characterized by different parameter-pairs $(C,D)$ not used for training. These predictions are visually indistinguishable from those obtained with the physics-based model $\mathbf g(\mathbf m)$ described in section~\ref{sec:fracture_network}.

\begin{table}[htbp]
    \caption{The best-trial NN architecture consists of six hidden layers, $FC_i$ ($i=1,\dots,6$), with the corresponding weight matrix $\mathbf{W}_i$ and layer output $\mathbf s_i$ ($i=1,\dots,6$) in~\eqref{eqa:fc_net}. Bias parameters are added to each layer, but not shown in this table. }
    \label{tab:NN_arch}
    \centering
    \begin{tabular}{lll}
    \hline
    Layer & Weights & Layer output \\
    \hline
    Input & - & 6 \\
    $FC_1$& $\mathbf{W}_1: 256 \times 6$ & $\mathbf s_1: 256$ \\
    $FC_2$& $\mathbf{W}_2: 64 \times 256$ & $\mathbf s_2: 64$ \\
    $FC_3$& $\mathbf{W}_3: 512 \times 64$ & $\mathbf s_3: 512$ \\
    $FC_4$& $\mathbf{W}_4: 256 \times 512$ & $\mathbf s_4: 256$ \\
    $FC_5$& $\mathbf{W}_5: 32 \times 256$ & $\mathbf s_5: 32$ \\
    $FC_6$& $\mathbf{W}_6: 128 \times 32$ & $\mathbf s_6: 128$ \\
    Output & $\mathbf{W}_7: 50 \times 128$ & $50$ \\
    \hline
    \end{tabular}
\end{table}

\subsection{Bayesian inversion without prior information}
\label{sec:standard}

\begin{figure}[htbp]
    \centering
    \includegraphics[width=0.9\textwidth]{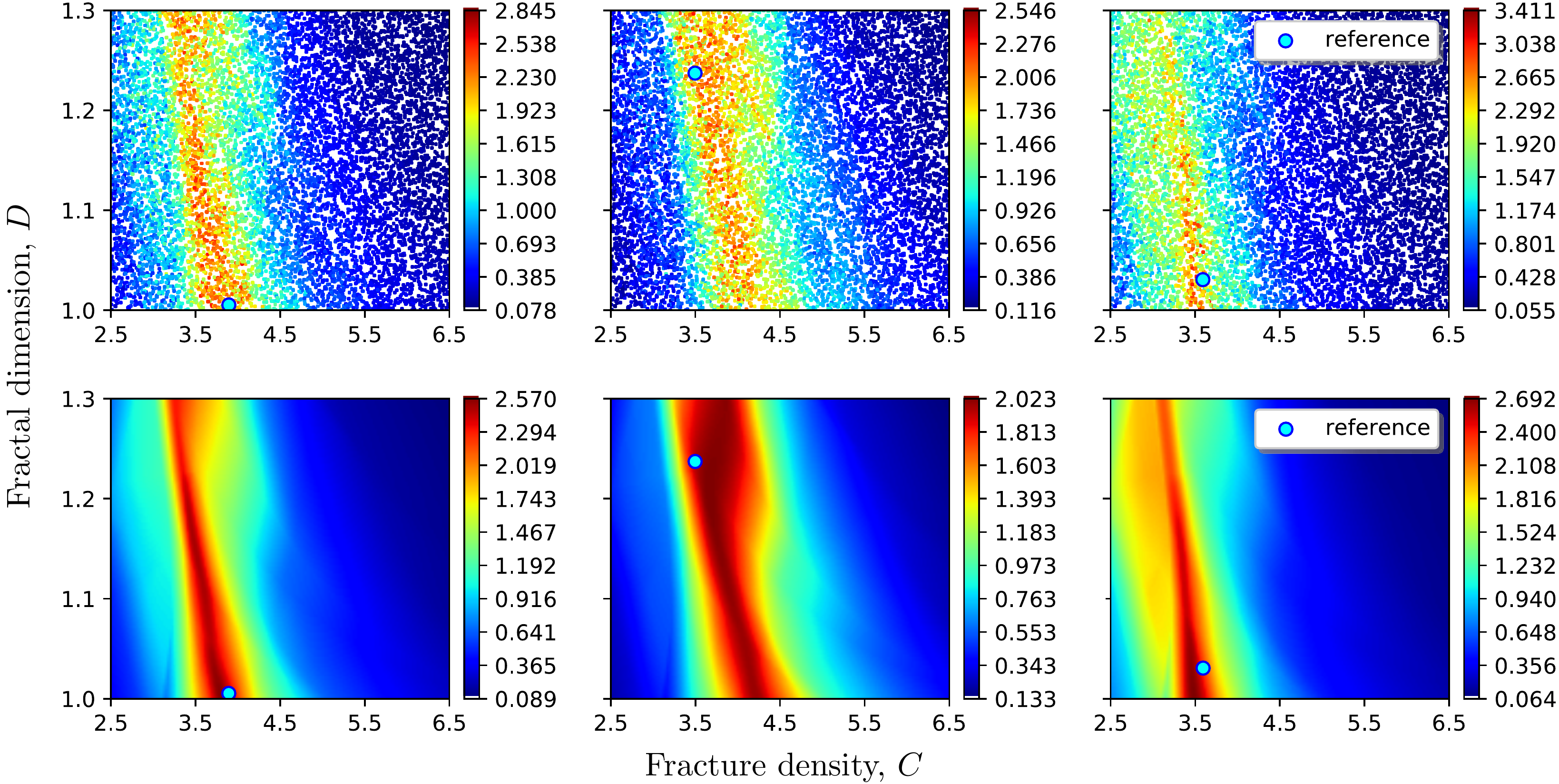}
    \caption{Examples of posterior PDFs of the DFN parameters $C$ and $D$, for three experiments defined by the reference parameter values (blue circles). These PDFs are computed via Bayesian assimilation of either $10^4$ runs of the physics-based model (top row) or additional $10^7$ runs of the FCNN surrogate (bottom row). }
    \label{fig:posterior1}
\end{figure}

We start with the Bayesian data assimilation and parameter estimation from section~\ref{sec:inv_strat}. Taking the uniform prior, $\gamma=0$ in~\eqref{eq:final_prior}, and assimilating the $N_\text{sim} = 10^4$ candidates provided by the physics-based model $\mathbf g$, this procedure yields the posterior PDFs of $C$ and $D$ shown in Figure~\ref{fig:posterior1}. While this non-informative prior indicates that all values of the parameters $(C,D)$ are equally likely, the sharpened posterior correctly assigns higher probability to the region containing the reference $(C,D)$ values. The relatively small number ($N_\text{sim} = 10^4$) of the forward solves of the physics-based model $\mathbf g$ manifests itself in granularity of the posterior PDF maps.

Significantly more forward model runs are needed to further sharpen these posterior PDFs around the true values of $(C,D)$ and to reduce the image pixelation. Generating the significant amounts of such data with the physics-based model is computationally prohibitive. Instead, we use $10^7$ additional candidates, corresponding to a $10^4\times 10^3$ mesh of the parameter space, provided by the FCNN surrogate. Figure~\ref{fig:posterior1} demonstrates that assimilation of these data (forward runs of the cheap FCNN surrogate) further reduces the band containing the unknown model parameters $(C,D)$ with high probability. %These distributions are improved as illustrated in Figure~\ref{fig:inverse_density} by using  data points, , and resulting in finer distributions that are consistent with the initial distributions shown in Figure~\ref{fig:inverse_density_sparse}. 
Generation of such large data sets with the physics-based model is four orders of magnitude more expensive than that with the FCNN. %The availability of a NN surrogates makes a difference between being able to solve this inverse problem or not.

%\begin{figure}[H]
%    \centering
%    \includegraphics[width=0.9\textwidth]{Figures/density_phy.pdf}
%    \caption{Examples of posterior distributions using an uniform prior ($\gamma=0$) obtained with $10^7$ data points provided by the $\mathbf{NN_f}$ model for 6 different pairs of parameter $(C,D)$ where the reference parameter value is shown with a blue circle for each example.}
%    \label{fig:inverse_density}
%\end{figure}

%In addition to improve the quality of the posterior distributions, the computational times reported in Table~\ref{tab:comp_cost} show that using the surrogate $\mathbf{NN_f}$ model results in reducing by a factor 4 the inverse analysis time, which is not feasible with the physically-based $\mathbf{PB_f}$ model.

\begin{table}[htbp]
\caption{Computational cost of the Bayesian inversion using the physics-based model $\mathbf g(\mathbf m)$ or the FCNN surrogate $\mathbf{NN(\mathbf m)}$. Each inversion requires $N_\text{sim}$ forward runs and takes time $T_\text{tot}$. The latter comprises time to train the model ($T_\text{train}$), time to execute the forward runs ($T_\text{run}$) and time to define the posterior PDF on the discretized parameter grid ($T_\text{grid}$). The running time for $\mathbf g(\mathbf m)$ is a projection based on the simulation time of $6560$~seconds that was necessary to run $10^4$ simulations. The FCNN was trained and executed on GPUs provided by GoogleColab. All times are in seconds.}
\label{tab:comp_cost}
\centering
\begin{tabular}{l l l l l l}
\hline
   & $N_\text{sim}$ & $T_\text{train}$ & $T_\text{run}$ & $T_\text{grid}$ & $T_\text{tot}$ \\
\hline
  $\mathbf g(\mathbf m)$   & $2 \times 10^8$ & $0$ & $1.312 \cdot 10^8$ & $5.47$ & $1.312 \cdot 10^8$ \\
  $\mathbf{NN}(\mathbf m)$  & $10^7$ & $37340$ & $1.26$ &  $5.47$ & $3.735 \cdot 10^4$ \\
\hline
\end{tabular}
\end{table}

The posterior PDFs displayed in Figure~\ref{fig:posterior1} show that the fracture density $C$ is well constrained and amenable to our Bayesian inversion, whereas the inference of the fractal dimension $D$ is more elusive. Examples of the DFNs in this study are provided in Figure~2 of~\cite{gisladottir2016particle}. They suggest that, for the parameter ranges considered, $C$ impacts the spatial extent of a fracture network, while $D$ affects the fracture-length distribution. Consequently, $C$ has a more significant impact on the overall structures.

\subsection{Bayesian inversion with data-informed priors}
\label{sec:modified}

\begin{figure}[htbp]
    \centering
    \includegraphics[width=0.9\textwidth]{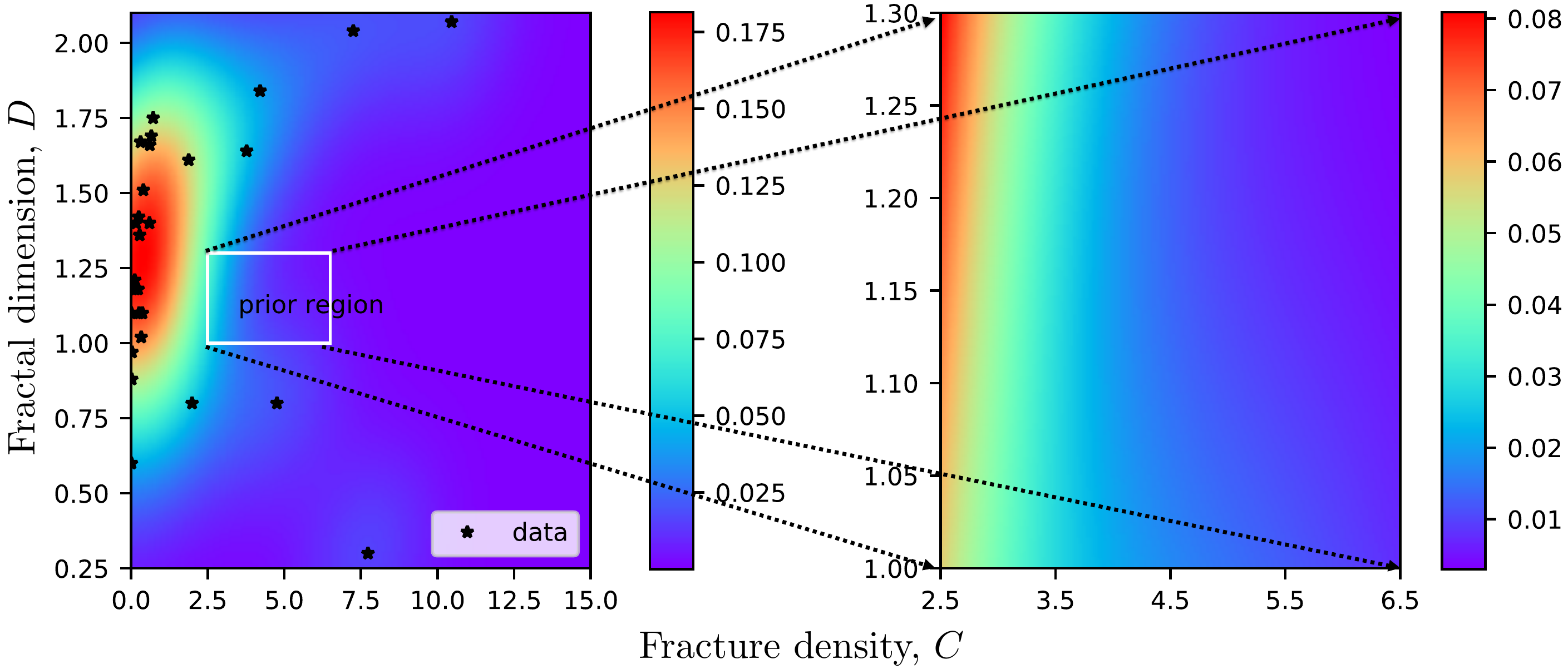}
    \caption{Prior joint PDF of $C$ and $D$ inferred from the field-scale data in \ref{app:corr_param} (left) and its rescaled counterpart over the parameter range used in our study (right).}
    \label{fig:prior}
\end{figure}

To refine the inference of parameters $C$ and $D$ from the breakthrough-time CDFs, we add some prior information. First, we observe that the field data reported in \ref{app:corr_param} suggest that $C$ and $D$ are correlated. These data are fitted with a shallow feed-forward NN resulting in the prior PDF of $C$ and $D$ shown in Figure~\ref{fig:prior}. These data vary over larger ranges than those used for $C$ and $D$ in the previous section; at the same time, most values correspond to $C<2$. That is because the field data come from a large number of different sites and from direct outcrop observations. Figure~9 in \cite{watanabe1995fractal} shows that a network with $C<2$ would have low connectivity. On the other hand, a DFN with a large $D$ is very dense, requiring large computational times to simulate and, possibly, being amenable to a (stochastic) continuum representation. Driven by these practical considerations, and to ascertain the value of this additional information, we restrict the prior PDF from Figure~\ref{fig:prior} to the same range of parameters as that used in the previous section. 

\begin{figure}[htbp]
    \centering
    \includegraphics[width=0.9\textwidth]{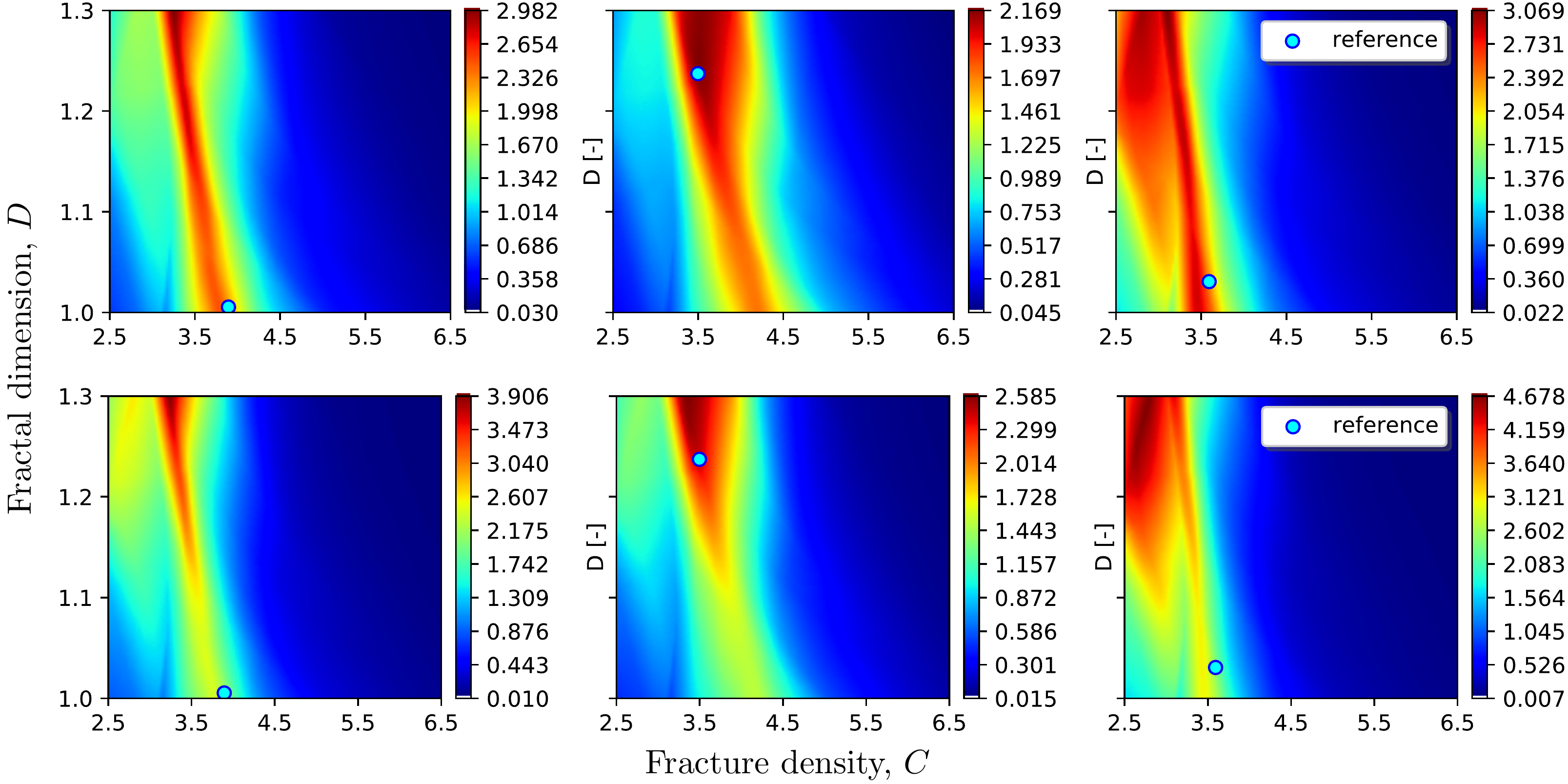}
    \caption{Examples of posterior PDFs of the DFN parameters $C$ and $D$ in the presence of prior information, for three experiments defined by the reference parameter values (blue circles). These PDFs are computed via Bayesian assimilation with the informative prior (Figure~\ref{fig:prior}), whose relative importance increases from $\gamma = 0.5$ (top) to $\gamma = 1.0$ (bottom). }
    \label{fig:posterior2}
\end{figure}

The relative importance given to the prior information about the DFN properties $C$ and $D$ (Figure~\ref{fig:prior}) is controlled by the parameter $\gamma$ in~\eqref{eq:gamma}. Large values of $\gamma$ correspond to higher confidence in the quality and relevance of the data reported in \ref{app:corr_param}. Figure~\ref{fig:posterior2} exhibits posterior PDFs of $C$ and $D$ computed via our Bayesian assimilation procedure with $\gamma = 0.5$ and 1. Visual comparison of Figures~\ref{fig:posterior1} and~\ref{fig:posterior2} reveals that the incorporation of the prior information about generic (not site-specific) correlations between $C$ and $D$ sharpens our estimation of these parameters, i.e., decreases the area in the parameter space where they are predicted to lie with high probability. Putting more trust in the prior, i.e., using a higher value of $\gamma$, amplifies this trend. However, the increase in certainty might be misplaced, as witnessed by several examples the reference parameter values fall outside the high probability regions.

%\begin{figure}[H]
%    \centering
%    \includegraphics[width=0.9\textwidth]{Figures/posterior2}
%    \caption{Examples of posterior distribution using the prior distribution shown in Figure~\ref{fig:prior} with $\gamma=0.5$ for 6 different pairs of parameters $(C,D)$ where the reference parameter value is shown with a blue circle for each example.}
%    \label{fig:inverse_density_0.5}
%\end{figure}

\begin{figure}[htbp]
    \centering
    \includegraphics[width=0.7\textwidth]{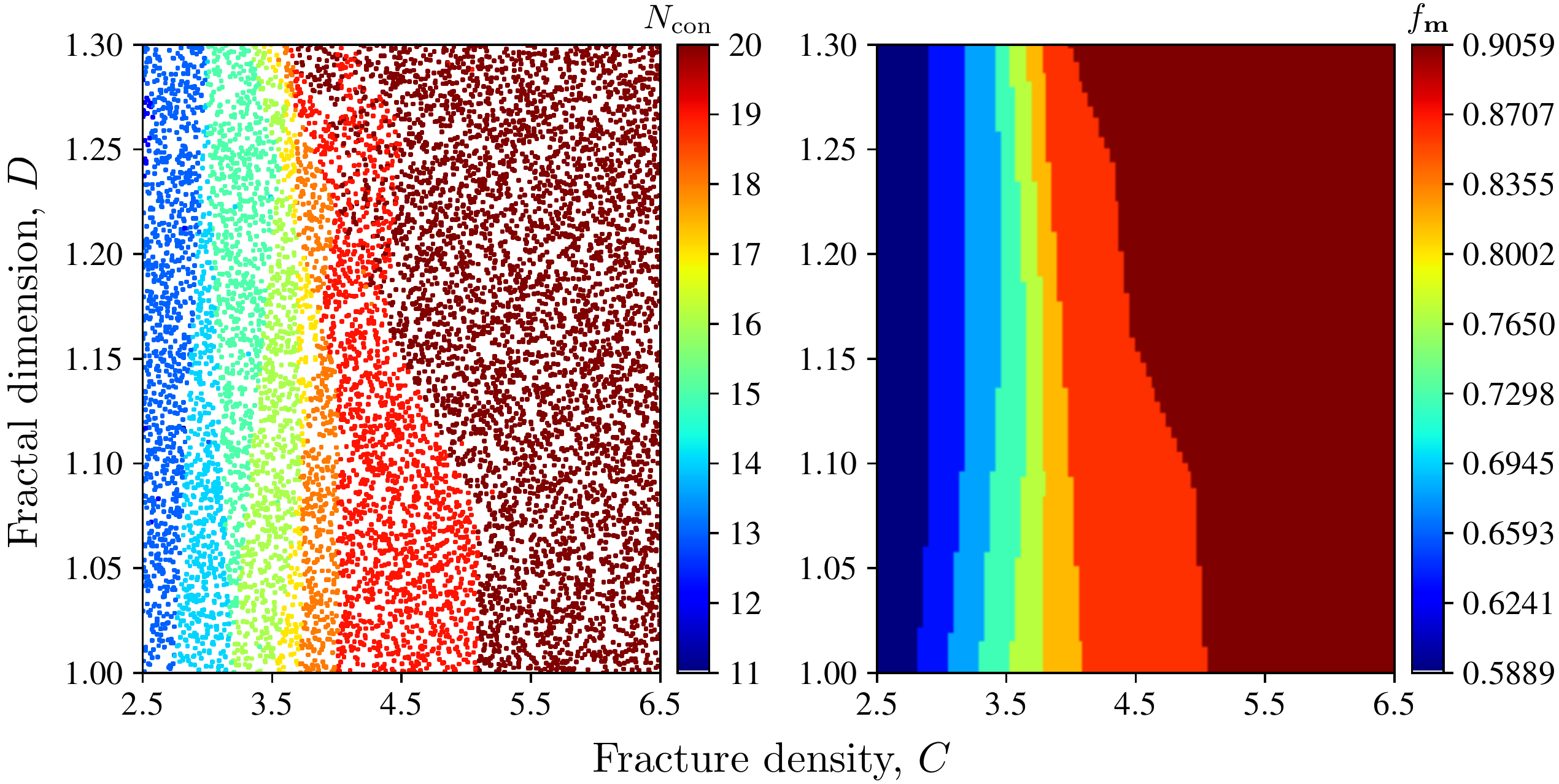}
    \caption{Number of connected networks, $N_\text{con}$, averaged over 20 random realizations of the DFN model with a given parameter pair $\mathbf m = (C,D)^\top$ (left); and corresponding prior PDF $f_{\mathbf m}$ in~\eqref{eq:prior_conn} (right).}
    \label{fig:conn_prior_density}
\end{figure}

Fracture network's connectivity is another potential source of information that can boost one's ability to infer the parameters $C$ and $D$ from CBTEs. Let $N_{\text{con}_i}$ denote the number of connected fracture networks among 20 random realizations of a DFN characterized by $(C,D)_i$. Figure~\ref{fig:conn_prior_density} exhibits $N_{\text{con}_i}$ for $N_\text{sim} = 10^4$ DFNs characterized by $(C,D)_i$ ($i=1,\dots,N_\text{sim}$), with the results interpolated to $10^4 \times 10^3$ mesh of the $(C,D)$ space by means of a shallow NN.  We define a prior PDF for $C$ and $D$ as 
\begin{equation}\label{eq:prior_conn}
    f_\textbf{m}(\tilde{\mathbf{m}}) \propto N^2_{\text{con}}(\tilde{\mathbf{m}}), \quad N_{\text{con}} \in [0,1,\dots, 20],
\end{equation}
which is properly normalized to ensure it integrates to one. This prior PDF, shown in Figure~\ref{fig:conn_prior_density}, assigns larger probability to those $(C,D)$ pairs that show higher connectivity in our data set. 

\begin{figure}[htbp]
    \centering
    \includegraphics[width=0.9\textwidth]{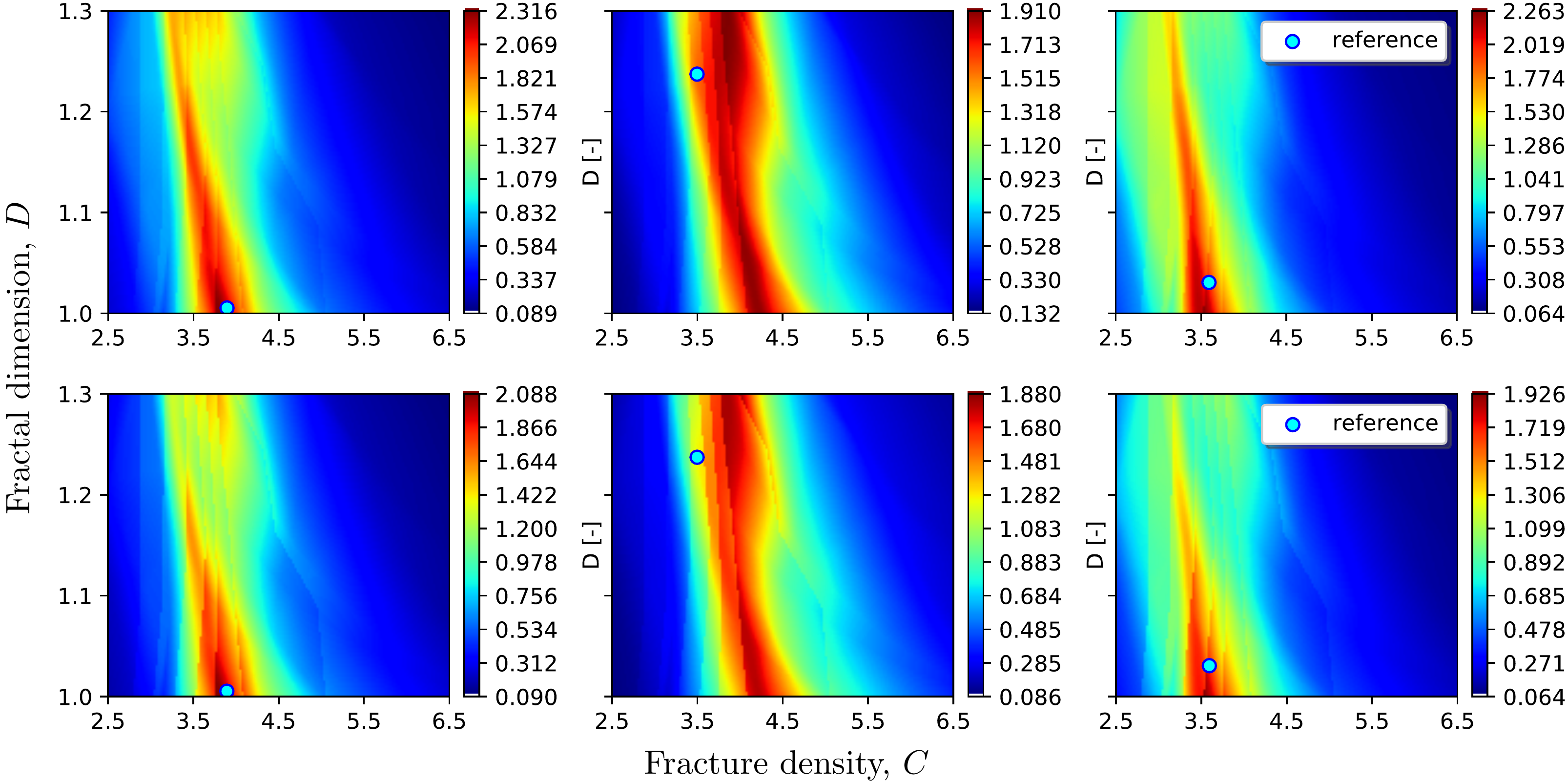}
    \caption{Examples of posterior PDFs of the DFN parameters $C$ and $D$ in the presence of prior information, for three experiments defined by the reference parameter values (blue circles). These PDFs are computed via Bayesian assimilation with the informative prior~\eqref{eq:prior_conn}, whose relative importance increases from $\gamma = 0.5$ (top) to $\gamma = 1.0$ (bottom).}
    \label{fig:post_conn_0_5}
\end{figure}

The Bayesian inference procedure with this prior yields the posterior joint PDFs of $C$ and $D$ in Figure~\ref{fig:post_conn_0_5}. These distributions are sharper than those computed with either  uninformative (Figure~\ref{fig:posterior1}) or correlation-based (Figure~\ref{fig:posterior2}) priors, indicating the further increased confidence in the method's predictions of $C$ and $D$.
%Figures~\ref{fig:inverse_density_0.5}, \ref{fig:inverse_density_1}, \ref{fig:post_conn_0_5} and \ref{fig:post_conn_1} show that adding prior information to our inverse analysis results in general in reducing the extent of the highest probability zones (dark red zones) in comparison with the posterior distributions that are computed without prior information and shown in Figure~\ref{fig:inverse_density}. 
As before, assigning more weight to the prior, i.e., increasing $\gamma$, reduces the area of the high-probability regions in the $(C,D)$ space. This increased confidence in predictions of $C$ and $D$ is more pronounced when the connectivity-based prior, rather than the correlation-based prior, is used. %even more important decrease of the high probability zones since the extent of the red zones decreases from Figures~\ref{fig:inverse_density_0.5} to \ref{fig:inverse_density_1} and from Figures~\ref{fig:post_conn_0_5} to \ref{fig:post_conn_1}. 
The connectivity information also ensures that this confidence is not misplaced, i.e., the reference parameter values lie within the high-probability regions.  % location of the reference parameter value is better with the prior defined from the connectivity of the system as it is still located in the highest probability zones for most of the cases shown in Figures~\ref{fig:post_conn_0_5} and \ref{fig:post_conn_1}.

%\begin{figure}[htbp]
%    \centering
%    \includegraphics[width=0.9\textwidth]{Figures/density_post_1_conn}
%    \caption{Examples of posterior distribution using the prior distribution shown in Figure~\ref{fig:conn_prior_density} with $\gamma=1$ for 6 different pairs of parameters $(C,D)$ where the reference parameter value is shown with a blue circle for each example.}
%    \label{fig:post_conn_1}
%\end{figure}

%%%%%%%%%%%%%%%%%%%%%% IV. CONCLUSIONS %%%%%%%%%%%%%%%%%%%%%%%%%%%
\section{Conclusions}
\label{sec:conclusion}

We developed and applied a computationally efficient parameter-estimation me\-thod, which makes it possible to infer the statistical properties of a fracture network from cross-borehole thermal experiments (CBTEs). A key component of our method is the construction of a neural network surrogate of the physics-based model of fluid flow and heat transfer in fractured rocks. The negligible computational cost of this surrogate allows for the deployment of a straightforward grid search in the parameter space spanned by fracture density $C$ and fractal dimension $D$. %algorithm method to characterize the fracture properties $(C, D)$ given the particle arrival time CDF measurement. A neural network is trained with a $\{(C,D)_i, ICDF_i\}_{i=1}^{N_d=10000}$ dataset to replace the forward transport model for computational efficiency. The likelihood of a given pair of $(C,D)$ is computed by comparing the output result of the neural network surrogate model with the ICDF measurement. This likelihood, together with a uniform or nonuniform prior density of $(C,D)$, determines the Bayesian-like posterior density of $(C,D)$. The posterior distribution on the domain $(C,D)\in [2.5, 6.5]\times[1.0, 1.3]$ is obtained on very fine $10000 \times 1000$ mesh grids with the method described above. 
Our numerical experiments lead to the following major conclusions.
\begin{enumerate}
\item The neural network surrogate provides accurate estimates of an average inverse cumulative distribution function (iCDF) of breakthrough times, for the fracture network characterized by  given parameters $(C,D)$.

\item In the absence of any expert knowledge about $C$ and $D$, i.e., when an uninformative prior is used, our method---with the likelihood function defined in terms of the Hellinger distance between the predicted and observed iCDFs---significantly sharpens this prior, correctly identifying parameter regions wherein the true values of $(C,D)$ lie.

\item Incorporation of the prior information about generic (not site-specific) correlations between $C$ and $D$ sharpens our estimation of these parameters, i.e., decreases the area in the parameter space where they are predicted to lie with high probability. Putting more trust in the prior, i.e., using a higher value of $\gamma$, amplifies this trend. However, the increase in certainty might be misplaced, as witnessed by several examples the reference parameter values fall outside the high probability regions.

\item Incorporation of the prior information about a fracture network's connectivity yields the posterior joint PDFs of $C$ and $D$ that are sharper than those computed with either  uninformative or correlation-based priors, indicating the further increased confidence in the method's predictions of $C$ and $D$.

\item The increased confidence in predictions of $C$ and $D$ is more pronounced when the connectivity-based prior, rather than the correlation-based prior, is used. The connectivity information also ensures that this confidence is not misplaced, i.e., the reference parameter values lie within the high-probability regions.
\end{enumerate}

% %%%%%%%%%%%%%%%%%%
% Figure~\ref{fig:param_corr} shows the correlated values of parameters $C$ and $D$ that are used in expression~\eqref{eq:nb_fract}. 
% \begin{figure}[H]
% \centering
% \includegraphics[width=0.7\textwidth]{Figures/param_corr.pdf}
% \caption{Correlation between parameters $C$ and $D$ from the values reported in Table~\ref{tab:appA1}}
% \end{figure}

%\newpage
\appendix
\section{Field-scale characterization of fracture networks}\label{app:corr_param}

For the sake of completeness, we report in Table~\ref{tab:appA1} the field-scale observations of fracture networks from~\cite{Bonnet2001}. These are accompanied by our calculation of the corresponding values of parameters $C$ and $D$ in the WT model of fracture networks.

{\small
\noindent
Table A.1: Fracture number ($N_\text{f}$), power-law exponent ($a$), surface area ($S$), minimum fracture length ($l_\text{min}$), and density parameter $\alpha$ for various fracture networks reported in Table 2 in \cite{Bonnet2001}. The corresponding values of fracture density ($C$) and fractal dimension ($D$) in the WT network model~\eqref{eq:nb_fract} are determined from the parameter relationships in Section~\ref{sec:fracture_network}.
\begin{longtable}{lllllll}%[htbp]
% \caption{}
\label{tab:appA1}\\
% \begin{tabular}{lllllll}
\hline
    $N_\text{f}$  [-]	& $a$  [-]	& $S$ [m$^2$]	& $l_\text{min}$ [m]	& $\alpha$  [-]	& $D$  [-]	& $C$  [-] \\ \hline
    107	& 1.74	& 24 & 0.1 & 0.60035 & 0.74	& 86.80731 \\
    121	& 2.11	& 25	& 0.1	& 0.41703	& 1.11	& 45.46014 \\
    3499 & 1.88 & 2.70$\cdot 10^{11}$ & $10^{3}$ &	4.97809$\cdot 10^{-6}$ &	0.88 &	0.01979\\
    120	& 0.9&	8.25$\cdot 10^{7}$&	40&	-1.00582$\cdot 10^{-7}$&	-0.1&	0.00012\\
    101&	1&	2.62$\cdot 10^{7}$&	57&	0&	0&	NaN\\
    300	&1.76& NP	&	7.00$\cdot 10^{3}$&	NaN	&0.76	&NaN\\
    380&	1.9	&3.43$\cdot 10^{3}$&	3	&0.26777&	0.9&	113.05832\\
    350	&2.1&	1.26$\cdot 10^{8}$&	220	&0.00115&	1.1	&0.36680\\
    1000&	3.2&	1.60$\cdot 10^{9}$&	380	&0.65137&	2.2	&296.07649\\
    1000&	2.1&	1.65$\cdot 10^{10}$&	2.00$\cdot 10^{3}$&	0.00028&	1.1&	0.25921\\
    800&	2.2&	2.50$\cdot 10^{1}$&	6.00$\cdot 10^{-2}$&	1.31254&	1.2&	875.02702\\
    380&	2.1	&NP&	2.50$\cdot 10^{3}$&	NaN&	1.1&	NaN\\
    1700&	2.02&	1.00$\cdot 10^{10}$&	1.00$\cdot 10^{3}$&	0.0002&	1.02&	0.33182\\
    260&	1.3&	8.75$\cdot 10^{3}$&	1.00&	0.00891&	0.3	&7.72571\\
    100&	1.8	&2.10$\cdot 10^{3}$&	1.00&	0.03809	&0.8	&4.76190\\
    873&	2.64&	3.40$\cdot 10^{1}$&	5.00$\cdot 10^{-3}$&	0.00709	&1.64&	3.7745\\
    320&	2.61&	2.07$\cdot 10^{7}$&	4.00$\cdot 10$&	0.00945&	1.61&	1.87779\\
    50&	1.67	&2.90$\cdot 10^{7}$&	7.00$\cdot 10$&	1.99004$\cdot 10^{-5}$&	0.67&	0.00148\\
    180&	1.97&	2.80$\cdot 10^{8}$&	3.00$\cdot 10^{2}$&	0.00016&	0.97&	0.02925\\
    400&	2.21&	1.20$\cdot 10^{8}$&	4.00$\cdot 10$&	0.00035&	1.21&	0.11573\\
    250&	2.11&	2.50$\cdot 10^{11}$&	4.50$\cdot 10^{3}$&	1.26005$\cdot 10^{-5}$&	1.11&	0.00284\\
    400&	2.84&	2.90$\cdot 10^{11}$&	5.50$\cdot 10^{3}$&	0.01935&	1.84&	4.20716\\
    70&	2.67&	3.60$\cdot 10^{9}$&	1.60$\cdot 10^{3}$&	0.00728&	1.67&	0.30533\\
    150&	2.66&	5.10$\cdot 10^{9}$&	1.25$\cdot 10^{3}$&	0.00675&	1.66&	0.61021\\
    200&	3.07&	6.20$\cdot 10^{9}$&	1.00$\cdot 10^{3}$& 	0.10829&	2.07& 	10.46329\\
    1034& 	2.51&	8.70$\cdot 10^{7}$&	1.00$\cdot 10$& 	0.00058& 	1.51& 	0.39767\\
    40 &	1.6 &	2.00$\cdot 10^{4}$& 	6.00$\cdot 10^{-2}$& 	0.00022 &	0.6& 	0.01479\\
    318 &	2.42&	1.69$\cdot 10^{8}$& 	7.00$\cdot 10$& 	0.00111 &	1.42&	0.24946\\
    291 &	2.69&	1.69$\cdot 10^{8}$ &	7.00$\cdot 10$& 	0.00382 &	1.69&	0.65783\\
    78 &	2.1&	1.69$\cdot 10^{8}$ &	1.00$\cdot 10^{2}$ &	8.04638$\cdot 10^{-5}$&	1.1& 	0.00570\\
    218 &	2.02 &	1.00 & 	2.00$\cdot 10^{-2}$ &	4.11251&	1.02 &	878.94881\\
    111 &	3.04 &	8.40$\cdot 10^{7}$& 	2.00$\cdot 10^{2}$ &	0.13328& 	2.04& 	7.25217\\
    470 &	1.8 &	1.17$\cdot 10^{4}$&	6.00$\cdot 10^{-2}$&	0.00338&	0.8 &	1.98852\\
    417&	2.18 &	6.00$\cdot 10^{7}$ &	4.00$\cdot 10$ &	0.00064&	1.18 &	0.22519\\
    201 &	2.4 &	3.00E-01 &	1.50E-04&      	0.00416	&1.4&	0.59676\\
    100	&2.4&	6.00$\cdot 10^{8}$&	7.00$\cdot 10^{2}$&	0.00224&	1.4&	0.16032\\
    1034&	2.36&	8.70$\cdot 10^{7}$&	1.00$\cdot 10$&	0.00037&	1.36&	0.28153\\
    450&	2.18&	2.20$\cdot 10^{8}$&	7.00$\cdot 10$&	0.00036&	1.18&	0.13843\\
    350&	2.75&	1.50$\cdot 10^{9}$&	1.80$\cdot 10^{2}$&	0.00361&	1.75&	0.72239\\
    300&	2.37&NP	&	1.00$\cdot 10^{2}$&	NaN&	1.37&	NaN\\
\hline
% \end{tabular}
\end{longtable}
}

\section*{Acknowledgments}

This project was facilitated by the generous grant from the France-Stanford Center at Stanford University. The work of ZZ and DT was supported in part the US Department of Energy GTO Award Number DE-EE3.1.8.1 ``Cloud Fusion of Big Data and Multi-Physics Models using Machine Learning for Discovery, Exploration and Development of Hidden Geothermal Resources'' and by a gift from Total. There are no data sharing issues since all of the numerical information is provided in the figures produced by solving the equations in the paper.

%%%%%%%%%%%%%%% REFERENCES %%%%%%%%%%%%%%%%%%%%%%%%%
%\newpage
%\bibliographystyle{elsarticle-num}
% \bibliographystyle{agufull08}
\bibliography{FracTherm}

\end{document}